\documentclass[letterpaper]{jpconf}
\usepackage{graphicx}
\usepackage{gensymb}
\newcommand{\sto}{SrTiO$_3$}
\newcommand{\lao}{LaAlO$_3$}
\begin{document}
\title{Domain walls in strontium titanate}

\author{C. Stephen Hellberg}

\address{U.S. Naval Research Laboratory, Washington, DC 20375, USA}

\ead{steve.hellberg@nrl.navy.mil}

\begin{abstract}
Density functional theory calculations of twin-domain walls in strontium titanate are presented.
The two possible domain walls are characterized.
The domain wall type is set by the relative phase of the octahedral rotations in the 
body-centered-tetragonal SrTiO$_3$ on either side of the domain wall.
The width of the domain walls is estimated to be approximately 3 nm.
\end{abstract}

\section{Introduction} 

\sto\ has the cubic perovskite crystal structure at room temperature.
Below approximately 105 K, it undergoes a cubic to tetragonal structural transition \cite{lytle64,sai00,howard05,zubko07,evarestov11,scott12,morozovska12,honig13}.
The oxygen octahedra rotate along one of the three crystallographic axes, and there 
is a slight elongation along that axis.
The rotations alternate in each layer, making the crystal structure body-centered tetragonal
%\cite{mehl17}.
\cite{setyawan10,mehl17}.
Twin-domain walls form between different tetragonal domains \cite{tagantsev01,merz16,gray16,schiaffino17}.
%zykova14 I think this may be incorrect--Octahedra tilt

At even lower temperatures,
\sto\ becomes a quantum paraelectric: ferroelectric ordering is suppressed by quantum fluctuations.  
When grown with a heavier isotope of oxygen, $^{18}$O, \sto\ becomes ferroelectric at 23 K \cite{itoh99}.
With the much more common $^{16}$O, \sto\ is paraelectric, but with local
ferroelectric regions \cite{tikhomirov02}.
Strain can  be used to induce bulk ferroelectricity in \sto\, by growing it on a substrate with a slightly
different lattice  constant \cite{pertsev00,aguirre04,haeni04,woicik06,woicik07,kourkoutis08,maitri09,hellberg12}.

In recent years, growth of oxide heterostructures with atomic-scale precision has become possible
\cite{ohtomo02,muller04}.
In particular, construction of interfaces with mismatched polarities, which is not possible in
semiconductor heterostructures \cite{harrison78}, was achieved in the \lao/\sto\ heterostructure
\cite{ohtomo04,thiel06,nakagawa06}.

\lao\ and \sto\ both have the same cubic perovskite crystal structure at room temperature,
but in \lao\ both cations have nominal charge $+3$, while Sr is $+2$ and Ti is $+4$.
Both materials are conventional band insulators, and they are nearly lattice matched \cite{pai18}.
%\cite{imada98,hellberg99kbs}. 
When grown in the [001] direction, each layer in \sto\ is nominally charge neutral, while the layers
in \lao\ are alternatively positive and negative.
The (001) surface can be prepared so that it is terminated by a TiO$_2$ layer \cite{ohtomo04}.
The \lao\ grown on this surface naturally forms a (largely) LaO layer on top of the TiO$_2$ layer \cite{maitri13}
forming an $n$-type interface \cite{nakagawa06}.
Once the \lao\ film exceeds a critical thickness of close to four unit cells (approximately 1.6 nm),
a conducting interface forms at the interface \cite{thiel06}.
The carriers are electrons in the conduction band of the \sto\ which are electrostatically bound to the interface.

At a thickness of approximately three unit cells, the nominally insulating interface can be made conducting
locally by patterning the \lao\ surface with a positively charge Atomic Force Microscope (AFM) tip \cite{cen08}.
The patterning is done at room temperature in humid air---it is believed that the AFM changes the surface
charge locally by removing adsorbed OH$^-$ molecules \cite{bi10}.
Nanostructures with dimensions as small as 2 nm can be patterned in this manner \cite{cen09}.
This technique has been used to pattern quantum dots in \sto\ \cite{cheng15}.
Tunneling through dots created in this manner 
showed that electrons pair in this system at temperatures several times
the superconducting critical temperature in \sto\ \cite{eagles69}.
In a separate experiment on wires of varying widths, it was found that the superconducting critical current
is {\em independent} of the wire width, indicating that the superconducting transport is along
the edges of the wires \cite{pai18b}.

The surface patterning to create the wires dopes electrons into the \sto\, causing the \sto\ to
expand by orienting the tetragonal domains normal to the surface \cite{honig13}.
If the tetragonal domains in the undoped regions neighboring the wires remains oriented in the plane,
domain walls will be created between doped $\hat{z}$-domains under the wire and undoped $\hat{x}$- or $\hat{y}$-domains next 
to the wires \cite{pai18b}.
Other experiments have shown enhanced conductivity along domain walls in \sto\ \cite{kalisky13}
and slightly increased superconducting $T_c$ at domain walls \cite{noad16}.
Enhanced conductivity along domain walls has been observed
in other materials, such as doped WO$_{3}$ \cite{aird98,aird98b,kim10,catalan12},
along grain boundaries in graphene \cite{lahiri10,phillips15,phillips17},
and in self organized stripes in the cuprates \cite{hellberg99stripe,rajasekaran18}.
Enhanced superconductivity has been observed in artificially layered metamaterials \cite{smolyaninova16}.

In this paper, we present Density Functional Theory (DFT) calculations of domain walls in \sto\ to examine 
their atomic structure  and to determine the interactions between domain walls.
% as a function of their separation.

\begin{figure}
\begin{center}
\includegraphics[width=6.5in]{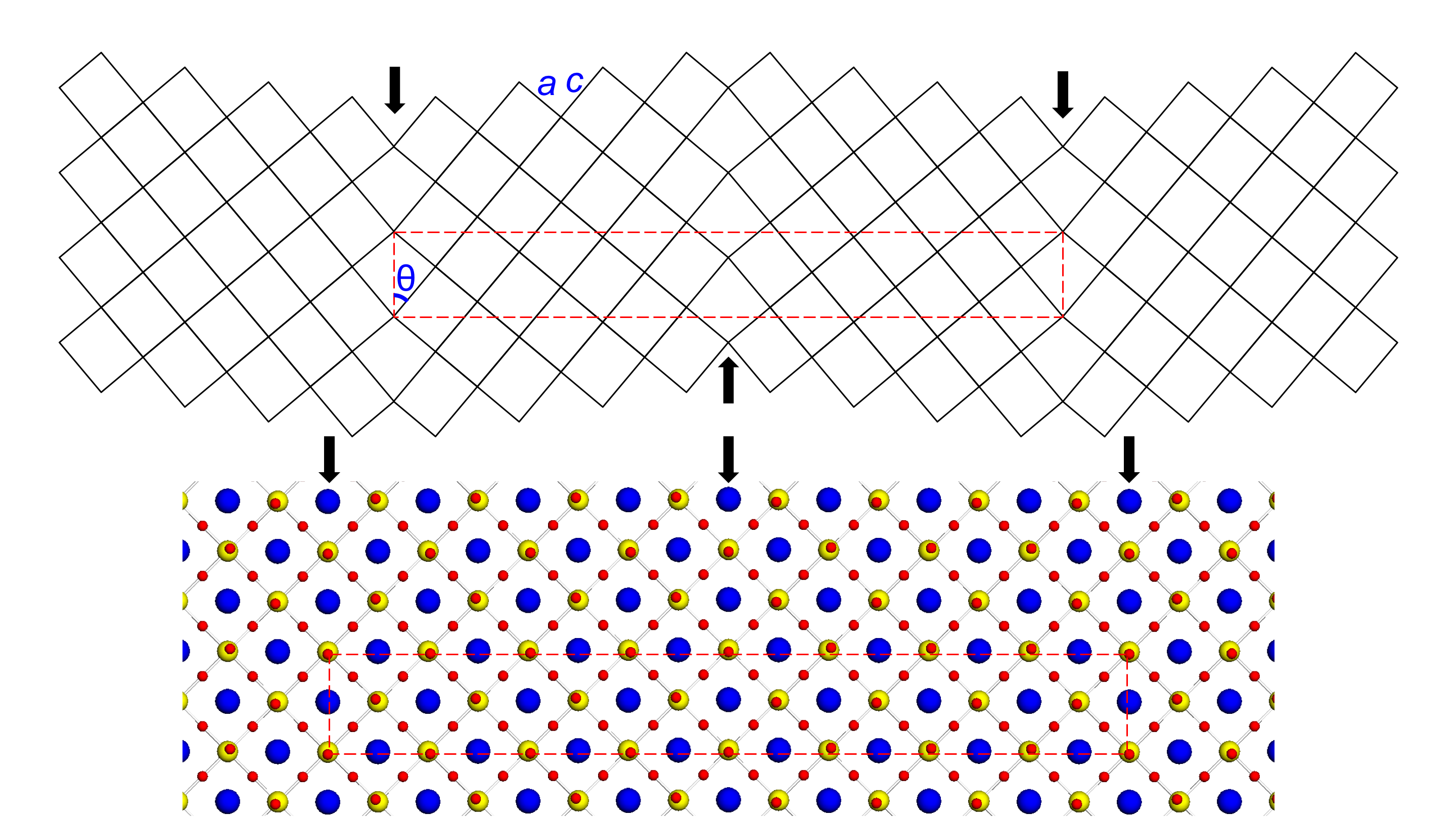}
\end{center}
\caption{\label{setup}
Top panel shows the schematic structure of the periodic computational cell containing
two domain walls outlined by the dashed red rectangle.
The $c/a$ ratio is increased for clarity.  Domain walls are indicated by the black arrows.
The $\hat{\bf x}$ direction is into the page, $\hat{\bf y}$ is to the top of the page, and $\hat{\bf z}$ is to the right.
The bottom panel shows the initial unrelaxed structure of a domain wall  with $N=8$ 
(see the text).  Sr atoms are blue, Ti are yellow, and oxygens are red.
This supercell is too small to support domain walls, and the structure relaxes to a uniform phase.
}
\end{figure}

\section{Computational details} 
%As discussed above, \sto\ has a body-centered-tetragonal structure at low temperatures.
%However, it is easier to begin the discussion by considering
%domain walls between regions with simply a tetragonal structure.
%
The properties of \sto\ and other perovskites depend sensitively on the choice of exchange-correlation
functional \cite{gebhardt17}.
We use the PBEsol functional, which has been shown to reproduce the lattice constants
of \sto\ very accurately \cite{perdew08,aschauer14}.
We use projector augmented wave functions as implemented in the Vienna {\em ab initio} simulation
package (VASP) \cite{kresse96,kresse99,blochl94}.
A plane wave cutoff of 283 eV and $4\times 4 \times 4$ Monkhorst-Pack $k$-points were used for
the 20-atom tetragonal cell.
A $4\times 4 \times 2$ $k$-point mesh was used for the domain wall calculations.
Atomic positions were relaxed until residual forces were less than 1 meV/\AA\@.
We find the optimal lattice constants of \sto\ are $a=3.893$ \AA\ and  $c=3.922$ \AA,
yielding a $c/a$ ratio of 1.0076.

Periodic computational cells are required for all calculations using VASP\@.
We construct the supercell for the domain wall calculations as shown in Fig.~\ref{setup}.
We rotate the tetragonal cell about the $\hat{x}$-axis by an angle $\theta = \tan^{-1}(a/c)$.
We tile the rotated cell along its elongated direction, and reflect the structure about
the $xy$-plane to create the other domain.
The resulting translation vectors for the supercell are:
\begin{eqnarray}
{\bf T}_1 & = &2 a \; \hat{\bf x}\nonumber\\
{\bf T}_2 & = &\sqrt{a^2+c^2} \;\hat{\bf y}\label{vecs}\\
{\bf T}_3 & = &2N\frac{ac}{\sqrt{a^2+c^2}} \;\hat{\bf z} \nonumber
\end{eqnarray}
for integer $N$, resulting in 4$N$ formula units in the supercell.
Note that the length of ${\bf T}_1$ must be an even multiple of $a$ to accommodate 
the octahedral rotations; there is no similar restriction on ${\bf T}_2$.

\begin{figure}
\begin{center}
\includegraphics[width=6.5in]{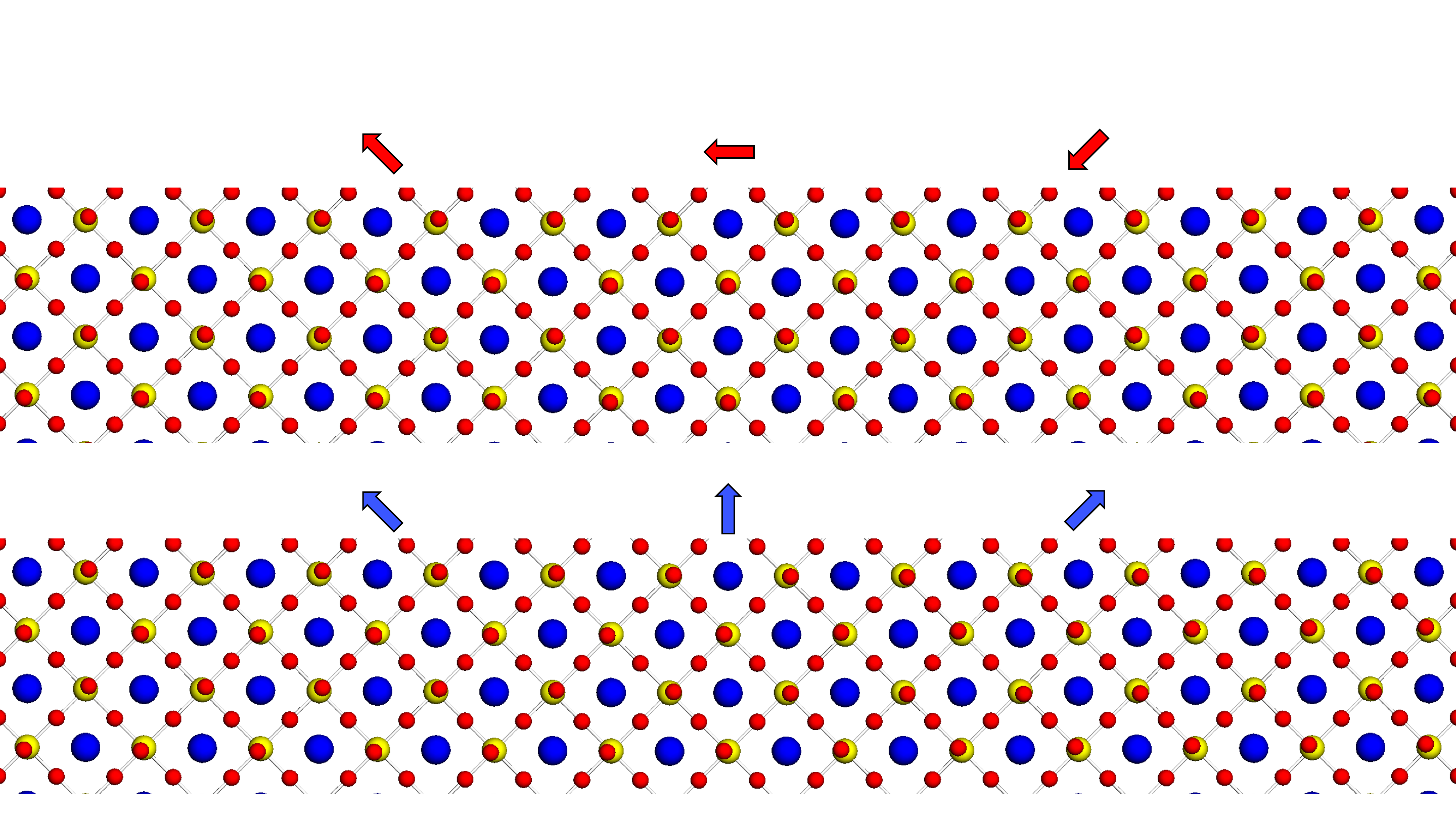}
\end{center}
\caption{\label{twodomainwalls}
Relaxed structure of isolated domain walls with $N=32$.
Each structure has a domain wall in the center of the figure.
The order parameter, the axis of rotation of the oxygen octahedra, is indicated by the arrows.
The upper structure has reflection symmetry about the domain wall;
the axis of octahedral rotation, indicated by the red arrows, rotates through the normal to the domain wall.
The lower structure has glide-reflection symmetry, with a glide vector of ${\bf G} = a \; \hat{\bf x}$ (into the page);
the axis of octahedral rotation, indicated by the blue arrows, rotates through the plane of the domain wall.
}
\end{figure}

\section{Atomic structure of domain walls} 

Due to the rotation of the octahedra, two types of domain wall are possible
as shown in Fig.~\ref{twodomainwalls}.
The first type has reflection symmetry about the domain wall; the second type is symmetric after
a reflection plus a glide vector of ${\bf G} = a \; \hat{\bf x}$.
The supercells for both domain walls have identical translation vectors.

We use the rotation axis of the oxygen octahedra as our order parameter.
In both types of domain wall, the order parameter gently rotates through the domain wall \cite{schiaffino17}.
With reflection symmetry, the order parameter rotates through the normal to the domain wall.
With reflection-glide symmetry, the order parameter rotates through the plane of the domain wall.
Notice that the type of domain wall is determined by
the relatives phases of the octahedral rotations in the tetragonal regions on either side of the domain wall.
Thus the type of domain wall is robust and cannot be altered by shifting the domain wall or by creating local
disorder.

%In both types of domain wall, the rotation of the octahedra gently tilts through the domain wall.
%With reflection symmetry, the rotation angle tilts through the normal to the domain wall.
%With reflection-glide symmetry, the rotation angle tilts through the plane of the domain.
%Notice that the type of domain wall is determined by
%the relatives phases of the octahedral rotations in the tetragonal regions on either side of the domain wall.
%Thus the type of domain wall is robust and cannot be changed by shifting the domain wall or by creating local
%disorder.
 
\begin{figure}
\begin{center}
\includegraphics[width=6.5in]{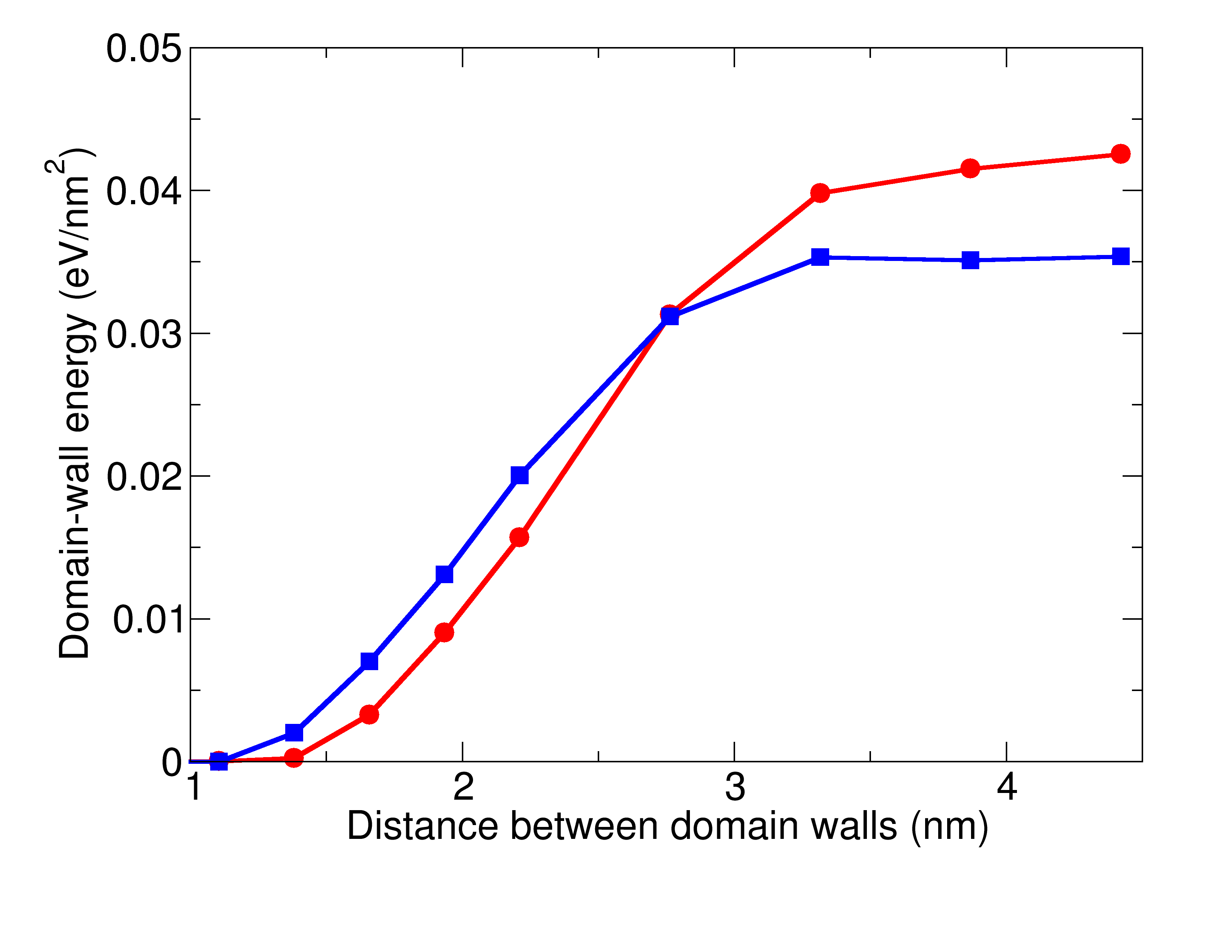}
\end{center}
\caption{\label{energy}Energy of domain walls as a function of their separation.
The domain-wall energy is computed from 
the energy difference between identical supercells containing four and two domain walls. 
The distance is defined as 25\% of the long axis of the supercell.
Domain walls with reflection symmetry are plotted as red circles; domain walls with glide-reflection symmetry
are blue squares.}
\end{figure}

\section{Interactions between domain wall} 

To study the interactions between domain walls, we performed a series of calculations varying the 
distance between the domain walls by varying $N$ in equation (\ref{vecs}).
To compute the formation energy for the domain walls, we need a reference energy, ideally of a pure
tetragonal phase with the same number of formula units.
However, tetragonal \sto\ cannot have the same translation vectors as the domain wall calculations.
We tried determining a reference energy from a calculation of the tetragonal structure with nearly the same
translation vectors as those used for the domain walls.
${\bf T}_1$ and ${\bf T}_1$ are unchanged and
${\bf T}_3$ is adjusted
by the shortest possible shift to accommodate the tetragonal structure:
\begin{equation}
\Delta{\bf T}_3  = N\frac{c^2-a^2}{\sqrt{a^2+c^2}} \;\hat{\bf y}.\label{shift}
\end{equation}
In this study, $N \leq 32$, and equation (\ref{shift}) is optimal; for significantly larger $N$, which would be computationally 
very expensive, equation (\ref{shift}) would no longer give the shortest shift consistent with a purely
tetragonal structure.
Unfortunately, the energy differences as a function of domain wall separation calculated in this manner were noisy,
probably due to the different $k$-point sampling in the differently shaped computational cells.

Computing the energy difference between a supercell containing
four and two domain walls allows exactly the same supercell to be used in both calculations.
The results are shown in Fig.~\ref{energy}.
In the limit of large separations, the two types of domain walls have comparable but clearly
different energies.
At distances less than 1 nm, the domain walls essentially annihilate, and the ions form a uniform phase,
resulting in no energy difference between the calculations with four and two domain walls.
These results show that interactions between domain walls in \sto\ does not extend beyond about 3nm.

\section{Summary}
Domain walls dominate the properties of many materials \cite{catalan12}.
In \sto\ experiments indicate that domain walls likely play a critical role in superconducting transport.
We computed the atomic structure of domain walls in \sto\ using density functional theory.
Due to the body-centered-tetragonal structure of \sto\ at low temperatures,
two types of domain walls are possible.
We find %both domain walls are metastable, and we show 
the static interactions between
domain walls are negligible beyond a distance of approximately 3 nm.

\ack Insightful discussions with Jeremy Levy, Massimiliano Stengel, Johnathan Ruhman, Patrick A. Lee, and Madeleine Phillips are gratefully acknowledged.  
This work was supported by the Office of the Secretary of Defense through the LUCI program and
by a grant of computer time from the DoD High Performance Computing Modernization Program.

%\bibliographystyle{../../nature}
%\bibliography{../../all}

\end{document}